\documentclass[]{aa}  

\usepackage{graphicx}
\usepackage{txfonts}

\usepackage{amsmath}

\usepackage[dvipsnames]{xcolor}
\usepackage{tikz}
\usetikzlibrary{positioning, calc, fit}

\usepackage{import}

\usepackage[breaklinks, colorlinks, citecolor=blue, linkcolor=blue]{hyperref}
\usepackage{soul} 

\usepackage{multirow}

\newcommand{\rsun}{$R_\odot$}

\newcommand{\panopticon}{\textsc{Panopticon}}
\newcommand{\platosim}{\texttt{PlatoSim}}

\begin{document}


\title{\panopticon{}: a novel deep learning model to detect single transit events with no prior data filtering in PLATO light curves}

\titlerunning{Transit detection in unfiltered PLATO light curves}


\author{
    H.~G. Vivien\inst{1} $^{\href{https://orcid.org/0000-0001-7239-6700}{\includegraphics[scale=0.01]{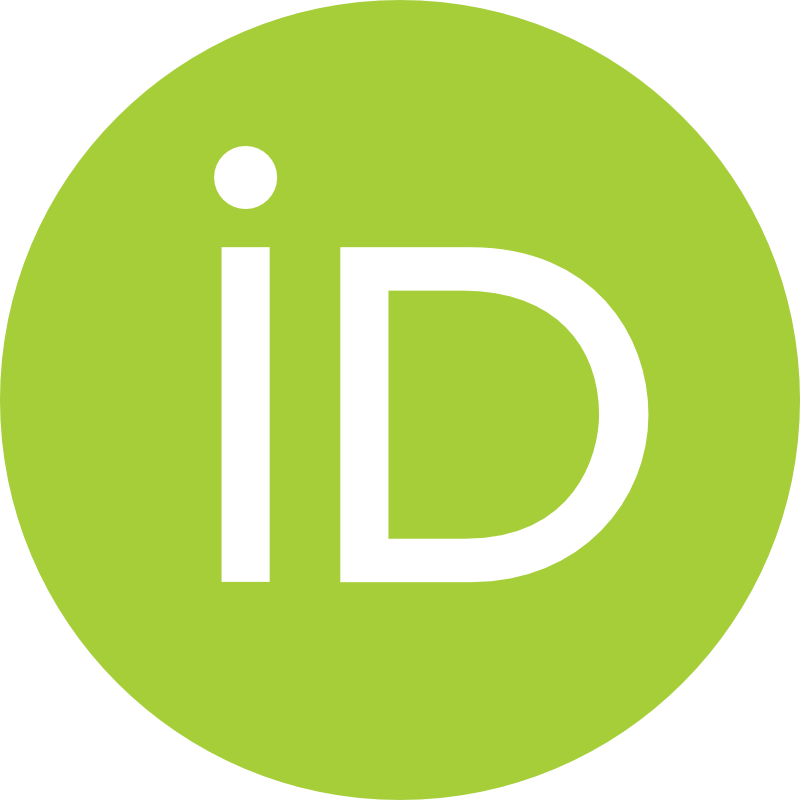}}}$ \and
    M. Deleuil\inst{1} $^{\href{https://orcid.org/0000-0001-6036-0225}{\includegraphics[scale=0.01]{fig/ORCID.png}}}$ \and
    N. Jannsen\inst{2} $^{\href{https://orcid.org/0000-0003-4670-9616}{\includegraphics[scale=0.01]{fig/ORCID.png}}}$ \and
    J. De Ridder\inst{2} $^{\href{https://orcid.org/0000-0001-6726-2863}{\includegraphics[scale=0.01]{fig/ORCID.png}}}$ \and
    D. Seynaeve\inst{2} $^{\href{https://orcid.org/0000-0002-0731-8893}{\includegraphics[scale=0.01]{fig/ORCID.png}}}$ \and
    M.-A. Carpine\inst{3} \and
    Y. Zerah\inst{1} $^{\href{https://orcid.org/0000-0003-1786-7367}{\includegraphics[scale=0.01]{fig/ORCID.png}}}$
}

\institute{
Aix Marseille Univ, CNRS, CNES, Institut Origines, LAM, Marseille, France\\ \email{hugo.vivien@lam.fr} \and
Institute for Astronomy, KU Leuven, Celestijnenlaan 200D bus 2401, 3001 Leuven, Belgium \and
AIM, CEA, CNRS, Université Paris-Saclay, Université Paris Diderot, Sorbonne Paris Cité, 91191 Gif-sur-Yvette, France
}

\authorrunning{Vivien et al.}

\date{Received Month dd, yyyy; accepted Month dd, yyyy}

 
\abstract
{}
{To prepare for the analyses of the future PLATO light curves, we develop a deep learning model, \panopticon{}, to detect transits in high precision photometric light curves. Since PLATO's main objective is the detection of temperate Earth-size planets around solar-type stars, the code is designed to detect individual transit events. The filtering step, required by conventional detection methods, can affect the transit, which could be an issue for long and shallow transits. To protect transit shape and depth, the code is also designed to work on unfiltered light curves.}
{The \panopticon{} model is based upon the Unet family architectures, able to more efficiently extract and combine features of various scale length, leading to a more robust detection scheme. We trained the model on a set of simulated PLATO light curves in which we injected, at pixel level, either planetary, eclipsing binary, or background eclipsing binary signals. We also include a variety of noises in our data, such as granulation, stellar spots or cosmic rays. We then assessed its capacity to detect transits in a separate dataset.}
{The approach is able to recover 90\% of our test population, including more than 25\% of the Earth-analogs, even in the unfiltered light curves. The model also recovers the transits irrespective of the orbital period, and is able to retrieve transits on a unique event basis. These figures are obtained when accepting a false alarm rate of 1\%. When keeping the false alarm rate low ($<0.01$\%), it is still able to recover more than 85\% of the transit signals. Any transit deeper than $\sim180$ppm is essentially guaranteed to be recovered.}
{This method is able to recover transits on a unique event basis, and does so with a low false alarm rate. Due to the nature of machine learning, the inference time is minimal; around 0.2\,s per light curve of 126\,720 points. Thanks to light curves being one-dimensional, model training is also fast, on the order of a few hours per model. This speed in training and inference, coupled to the recovery effectiveness and precision of the model make it an ideal tool to complement, or be used ahead of, classical approaches.}

\keywords{Planetary systems --- Planets and satellites: detection --- Techniques: photometry}

\maketitle


\section{Introduction}\label{sec:intro}

Out of the currently $\sim5700$ confirmed planets, close to 3900 have been discovered using transits\footnote{Data from \href{exoplanet.eu}{https://exoplanet.eu/}}. This approach was first successfully used to record the predicted transit of HD209458b in 1999 \citep{Charbonneau_2000}, and the first candidate detection came quickly after in 2002 \citep{Udalski_2002}, later confirmed in 2003 \citep{Konacki_2003}. Since then, space-based missions such as CoRoT \citep{AuvergeAl_2009} or Kepler/K2 \citep{Borucki_2010, Howell_2014} have been designed to acquire the photometry of multiple stars simultaneously, scaling up the ability to detect transits. Additionally, this is the only method allowing the direct measurement of certain physical parameters, such as planetary radius, when it is coupled to asteroseismology. To this effect, the second generation of space mission, TESS \citep{Ricker_2015} and CHEOPS \cite{Benz_2021}, now provide high-precision photometry of multiple targets.

However, even in high-precision photometry, stellar activity can prevent transit detection and proper characterization. This becomes even more of a problem for long period planets, where folding the light curve to increase the signal to noise ratio might not be an option. So far, the usual approach is to perform a periodicity analysis of the signal, for example with the box least square \citep[BLS;][]{KovacsAl_2002}, or the transit least square\footnote{TLS: \href{https://github.com/hippke/tls}{https://github.com/hippke/tls}} \citep[TLS;][]{HippkeHeller_2019} algorithms. For single transit event, it is mandatory to filter out the stellar activity perfectly to be able to assert the nature of the signal. Currently, using Gaussian processes to fit empirical models has proven effective, but quite costly in computation time. Because of the stochastic nature of Gaussian processes, it requires human supervision in order to avoid affecting the shape and depth of the transit.

Both the prior filtering and the search for periodic signal set stringent detection limits on the detectable population of exoplanets. For the forthcoming PLATO mission \citep{RauerAl_2014, RauerAl_2024}, whose prime goal is the detection and characterization of Earth analogs, this limit is going to become even more prevalent. Indeed, the mission's main challenge will be to detect unique transit events that are likely to be Earth-type, to ascertain the planetary nature of candidates, in order to begin follow-up campaigns as quickly as possible.

The recent and rapid rise of machine learning (ML) and its extension, deep learning (DL) methods for widespread data analysis, and its fast paced development, offer new opportunities both in applications and methodology. Its use in astrophysics has so far remained somewhat uncommon, despite proving effective at both identification and classification problems, even without prior data filtering. The exoplanet field appears well suited for ML/DL application, but surprisingly few studies have looked into the possible use cases. Most of the studies so far have focused on planet candidates vetting in Kepler/K2 \citep{MccauliffAl_2015, AnsdellAl_2018, ShallueVenderburg_2018, DattiloAl_2019}, Kepler and TESS \citep{ValizadeganAl_2022} or NGTS \citep{ArmstrongAl_2018}. Fewer still have looked at direct detection; \citet{ZuckerAl_2018}, which injected periodic signals in synthetic noisy light curves, and \citet{Malik_2022}, which investigated long period planets in TESS light curves.

In this work we present \panopticon{}, a DL model designed to detect single events in unfiltered light curves. Avoiding signal filtering prior to detection aims at preventing shallow transits from being phased out in fitted models (such as with Gaussian processes), therefore improving the detection of long-period planets. We train and test our approach on simulated PLATO light curves, and extract the position of likely threshold crossing events (TCE) in the light curves.

First we describe the architecture of the model in Sect.~\ref{sec:model} and our dataset in Sect.~\ref{sec:data_enviro}. We detail our results and performances in Sect.~\ref{sec:performances}. Finally, we present our conclusions in Sect.~\ref{sec:discussion}.

\section{Deep learning model}\label{sec:model}

In this paper, we present a DL model able to identify transit signals in a light curve. This is done by localizing the position of transit events. We do this in the context of the forthcoming ESA's PLATO mission, designed to determine the frequency of Earth-sized planets orbiting Sun-like stars. We opt for a classifier approach, rating the probability that a transit is occurring at each point of a light curve. This allows our approach to retrieve an arbitrary number of transits in a light curve, including the case of mono-transits. Additionally, we rely on the DL ability to extract and classify features to properly identify the events to bypass the filtering process.

\subsection{Architectures}\label{ssec:architecture}

We implement a custom 1-dimensional version of the Unet family: Unet, Unet++, Unet3+ \citep{RonnebergerAl_2015, ZhouAl_2018, HuangAl_2020}. This architecture is a type of fully convolutional neural network, that adds successive upsampling layers to the usual contracting networks. By combining the features extracted from the contracting during the upsamling process, the model can yield a high resolution output. For our light curves, the output generated acts as a one-to-one map of the input, where each point is classified individually based on neighboring context.

A Unet model can be seen as an auto-encoder with skip connections between the layers of the encoder part and the decoder part. The encoder extracts contextual information from the input, while the decoder builds the output point by point. During the encoding process the input is iteratively down-sampled, allowing a fixed-size kernel to extract information over a larger window at each step. Then, the decoder iteratively upsamples the output of the encoder, combining it with the features previously extracted at various timescales. The output of the decoder is therefore a segmentation map covering the input point-to-point, allowing for precise localization of the object of interest in the input light curve. Besides, a DL model with skip connections is beneficial, as they have been shown to increase training speed, and also allow for deeper network \citep{DrozdzalAl_2016}.

\begin{figure}
	\centering
	\includegraphics[width=\columnwidth]{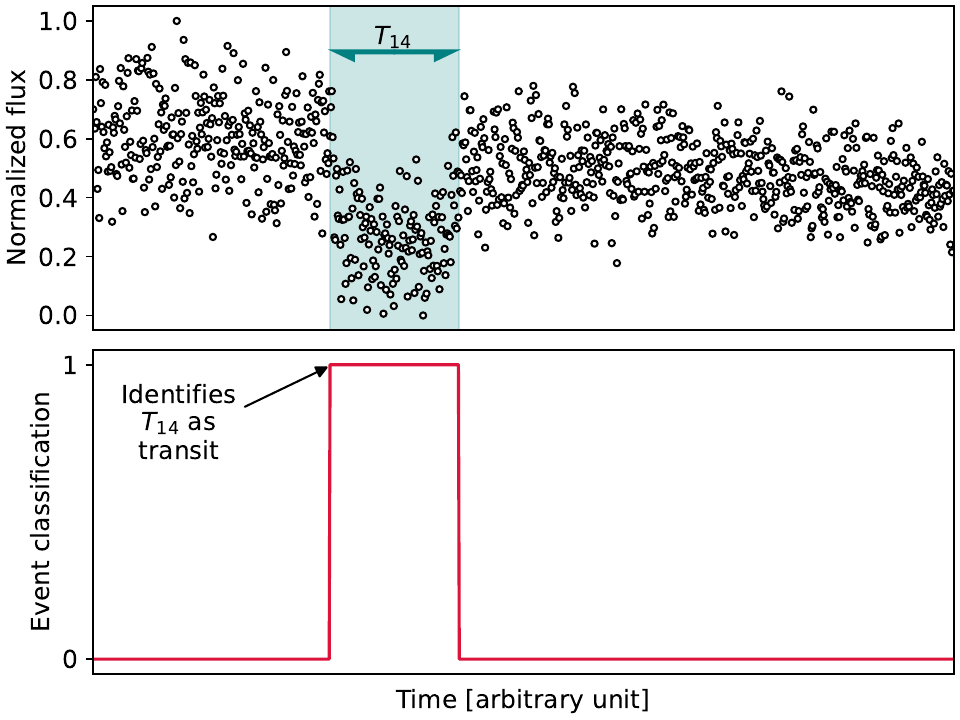}
	\caption{Theoretical input-output scheme of the model. A light curve, normalized between 0 and 1, is given as input to the model (top panel). We highlight the transit by the blue region. The model returns a classification map for the whole light curve (bottom panel).}
	\label{fig:theoretical_classification}
\end{figure}

Because this approach is a point-wise detection, it presents a few advantages. First and foremost, it allows any transit in a light curve to be detected individually, as points are classified based on local context. Second, we can extract the $T_0$ and duration for an arbitrary number of transits in a given light curve. Third, because the output yields a probability, it is possible to define the confidence level depending on the required certainty to extract the TCEs. Finally, because DL builds an internal noise model, there is no need for prior filtering of the light curves. Bypassing this step ensures that no shallow signal will be removed by mistake, and simplifies the detection process significantly. The theoretical detection of a transit in a light curve by our model is illustrated in Fig.~\ref{fig:theoretical_classification}, and the models architectures are given in Figs.~\ref{graph:unet}, \ref{graph:unetpp} and \ref{graph:unet3p}.

\subsection{Implementation}\label{ssec:implementation}

\tikzset{
    block/.style = {
        rectangle,
        draw = black!100,
        fill = black!10,
        minimum width = width("Convolution++"),
        rounded corners,
        node distance = 0.1cm
    },
    io/.style = {
        rectangle,
        draw = Mulberry!100,
        fill = Mulberry!10,
        rounded corners,
        node distance = 0.5cm
    },
    highlight/.style = {
        rectangle,
        draw = black!100,
        thick,
        rounded corners,
        node distance = 0.0cm
    },
    highlight_inv/.style = {
        rectangle,
        draw = black!0,
        line width = 0pt,
        rounded corners,
        node distance = 0.0cm
    },
    optional/.style = {
        rectangle,
        draw = teal!100,
        fill = teal!10,
        dashed,
        minimum width = width("Convolution+++"),
        rounded corners,
        node distance = 0.12cm
    },
    direct/.style = {
        arrows = {->[color=black]},
        draw = black,
        line width = 0.5mm
    }
}

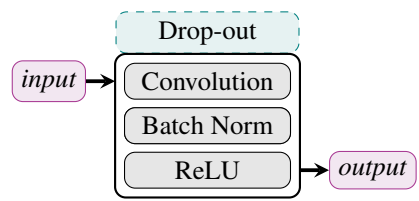
\begin{figure}
\centering
    \begin{tikzpicture}[x=1.5cm, y=1.5cm, >=stealth]

        \node[block](conv){Convolution};
        \node[highlight_inv, fit = (conv)](prac_conv){};
        \node[io](input)[left = of conv]{\textit{input}};
        \node[optional](drop)[above = of conv]{Drop-out};
        \node[block](norm)[below = of conv]{Batch Norm};
        \node[block](relu)[below = of norm]{ReLU};
        \node[highlight_inv, fit = (relu)](prac_relu){};
        \node[io](output)[right = of relu]{\textit{output}};
        \node[block](norm)[below = of conv]{Batch Norm};

        \node[highlight, fit = (conv) (relu)](base){};

        \draw[direct]   (input) edge (prac_conv)
                        (prac_relu) edge (output);
    
    \end{tikzpicture}
    
\caption{Basic convolution block used in the model. For Unet and Unet++ each node is made up of two consecutive occurrence of this block, but a single one is used for the nodes of Unet3+. The dropout layer is optional, and if used is applied only once per block.}
\label{graph:block}
\end{figure}

The three variants, Unet, Unet++ and Unet3+ are illustrated in Figs.~\ref{graph:unet}, \ref{graph:unetpp} and \ref{graph:unet3p}, respectively. The encoder and decoder can be seen as a series of nodes, each either extracting features or combining them, respectively. Each node is built upon a basic convolution block, illustrated in Fig.~\ref{graph:block}. This block consists of three base operations: a convolution, a batch normalization and a rectified linear unit. An additional, optional, drop-out layer can be included at the beginning of the block. The node of the backbone make use of either two consecutive blocks in the cases for Unet and Unet++, or a single block for Unet3+.

We identify the nodes of the model as $x^{i,j}$. The index $i$ corresponds to the depth of the node in the network. It increases with each downsample operation, and decreases with each upsample. Conversely, index $j$ tracks number of upsampling steps to reach a given node. We can therefore easily identify every node in the models. The backbone of the models corresponds to nodes $x^{i,0}$. Reciprocally, the decoder is made up of the nodes $x^{i,j>0}$. The backbone is common to all model and can be computed using:

\begin{equation}\label{eq:backbone}
    x^{i,0} = \begin{cases}
        \mathcal{N}\left( x^\mathit{input} \right), & i=0 \\ 
        \mathcal{N}\left(\mathcal{D}\left( x^{i-1,0} \right)\right), & i>0 \\ 
    \end{cases}
\end{equation}

\noindent where $\mathcal{N}$ is the operation assigned to the node using the default block, described above. $\mathcal{D}$ corresponds to the downsampling operation (Max Pooling; selecting the maximum value within a certain kernel). In this case, the kernel is set to a length of 2, and results in halving the resolution of the input at each level. Additionally, the number of feature channels also increases at each level. The decoders of each model can then be computed as follow:

\begin{equation}\label{eq:skip_connections}
    x^{i,j>0} = \begin{cases}
        \mathcal{N}\left(\left[ x^{i,0},~\mathcal{U}\left(x^{i+1,j-1}\right) \right]\right), & \text{Unet} \\
        \mathcal{N}\left(\left[ \left[x^{i,k}\right]_{k=0}^{j-1},~\mathcal{U}\left(x^{i+1,j-1}\right) \right]\right), & \text{Unet++} \\
        \mathcal{N}\left(\left[ \begin{aligned}
            \left[\mathcal{N}\left(\mathcal{D}\left(x^{k,0}\right)\right)\right]_{k=0}^{i-1}, \\
            \mathcal{N}\left(x^{i,0}\right), \\
            \left[\mathcal{N}\left(\mathcal{U}\left(x^{0,k}\right)\right)\right]_{k=0}^{j-1}
        \end{aligned}
        \right]\right), & \text{Unet3+}
    \end{cases}
\end{equation}

\noindent where $\mathcal{D}$ remains the downsampling operation, and $\mathcal{U}$ is the upsampling operation (Transpose Convolution for Unet and Unet++ or Upsample for Unet3+). The upsampling operation is setup so that it upsamples the data to the same length as the target node. Anything contained within $[]$ is concatenated feature-wise.

\begin{table}
	\centering
    \caption{Typical number of parameters for each architecture.}
	\begin{tabular}{ | r || c | c | c | }
	\hline
	Kernel & Unet & Unet++ & Unet3+ \\
	\hline\hline
	11 & 562\,713 & 651\,025 & 516\,385 \\
	31 & 1\,543\,353 & 1\,778\,865 & 1\,452\,225 \\
	61 & 3\,014\,313 & 3\,470\,625 & 2\,855\,985 \\
	\hline
	\end{tabular}
	\label{tab:param_number}
    \tablefoot{The kernel here refers to the number of learnable parameters $k_l$, and the number of initial feature channels is set to 4.}
\end{table}

The kernel size of the convolution operation is key on two aspects; (i) the number of trainable parameters and (ii) the coverage of the signal it offers. A larger kernel can increase the quality of the features identification, at the cost of longer training time. Additionally, the kernel must be large enough to encompass recognizable features within the signal. To achieve a good balance between feature quality, training time and feature coverage, we make substantial use of kernel dilation:

\begin{equation}\label{eq:kernel_length}
	k_s = k_t + (k_t - 1)(d - 1)
\end{equation}

\noindent where $k_s$ is the total length of the kernel, $k_t$ is the number of active parameters in the kernel, and $d$ is the dilation factor, namely, the spacing between active points in the kernel. For a default kernel where all active points are next to each other, the dilation factor is 1. This allows us to increase the size of the kernel for a fixed number of trainable parameters, at the cost of a lower resolution per feature. Typical kernel size for each architecture is shown in Table~\ref{tab:param_number}, for a constant depth of four and the number of initial feature maps set to eight. The number of learnable kernel parameters has a strong impact, while each model appears fairly similar. The Unet3+ version displays the smallest number of parameters for a given number of parameters, showing the advantage of the full skip connection over the nested skip connections of Unet++.

As described above, the goal of the model is to identify transit events directly within the stellar noise. We therefore first limit the model to a binary classification scheme, distinguishing two classes: "continuum" and "event". Each point in the light curve is assigned a likelihood score to belong to either the continuum (0) or an event (1). To retrieve the classification, we need to define a threshold to separate the two classes. Given the intrinsic class imbalance present within our data, due to the short nature of the transits compared with that of the light curve, it is not guaranteed that setting the threshold to 0.5 will yield the best results \citep{LiAl_2019}. To more effectively constrain the best threshold value, we evaluate the performance of the model for values ranging from 0.05 to 0.95.

Finally, to compare the performances of the model, we train multiple models over a wide range of parameters, leading to the comparison of multiple versions of the model. This allows to compare the aforementioned kernel sizes and coverage of the features. We limit the model to use a binary cross entropy (BCE) loss function, which has proven effective. We use the AdamW optimizer, with $\gamma=0.001$, $\beta_1=0.9$ and $\beta_2=0.999$. We set the feature dropout rate to 10\%.

\section{Dataset and environment}\label{sec:data_enviro}

\begin{figure*}
	\centering
	\includegraphics[width=\textwidth]{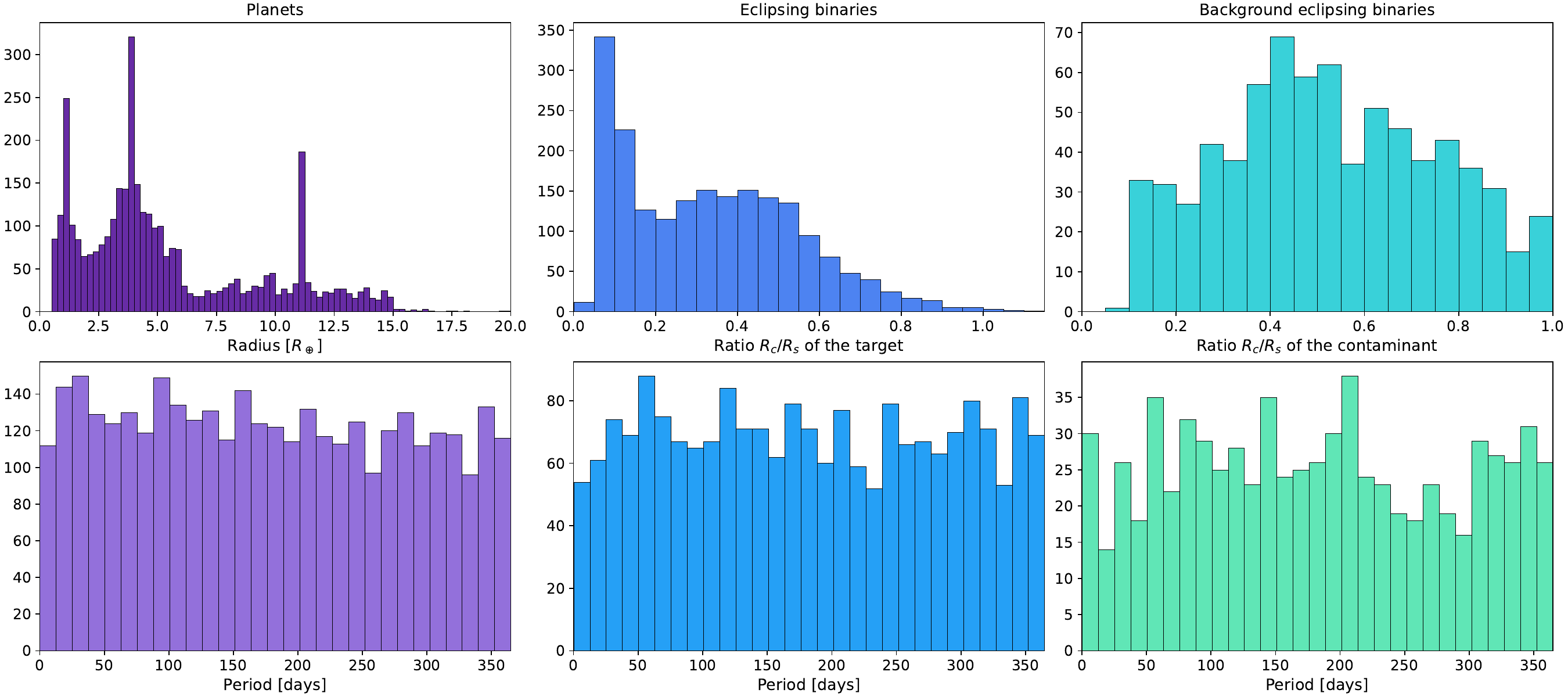}
	\caption{Histograms of the radii and periods of the bodies in our dataset. The left, center and right columns correspond respectively to the planets, the eclipsing binaries, and the background eclipsing binaries. The tree spikes in the planetary population correspond to an erroneous simulation run that didn't include the sampling of radii around the central values of the distribution.}
	\label{fig:data_hist}
\end{figure*}
 
To prepare a realistic dataset, tailored for the PLATO instrument, while controlling the astrophysics content of the light curves, we took advantage of the mission end-to-end camera simulator \platosim{} \citep{JannsenAl_2023}. \platosim{} is developed to generates accurate and realistic simulated images to be received from the PLATO satellite. It includes a wide range of instrumental noise sources at different levels: platform, camera, with realistic PSFs, and detector. With stars that can be observed by a variable number of cameras and the complexity of the instrument, which accommodates 26 cameras on a single optical bench, to control the observation conditions for a given star, we used \texttt{PlatoSim}'s toolkit, called \texttt{PLATOnium}, which, based on the PLATO input catalog \citep[PIC,][]{MontaltoAl_2021, NascimbeniAl_2022}, enables the simulator to be programmed in a friendly way. In addition to a realistic representation of the instrument, the other advantage of using \platosim, is that the signal is injected at pixel level. While this is not fundamental to test the mere detection, this is central for later identify our capacity to sort false positives generated by background eclipsing binaries, and bona fide planets.

To build our simulated dataset, we chose stars identified in the PIC as potential targets for the prime sample (P1; main sequence stars with $V_\mathrm{mag}<11$), as the signal-to-noise ratio (SNR) of these stars enables a detection of an Earth-like planet. We then simulate those stars, including various astrophysical signals:

\begin{description}
    \item[-] stellar activity effects that include granulation, stochastic oscillations and stellar spots
    \item[-] exoplanet transits, simulated with \texttt{BATMAN} \citep{Kreidberg_2015} 
    \item[-] eclipsing binaries, simulated with \texttt{ellc} \citep{Maxted_2016}
\end{description}

We can thus simulate a target combining all of these effects to produce either a transiting planet in front of an active star, or an eclipsing binary on the target, or an eclipsing binary on a nearby contaminant. All the physical characteristics used to generate the signal (masses, radii, effective temperature, orbital period, ephemerides, eccentricity, rotation period of the star, pulsation frequency, etc.) are also documented and saved.

The prime sample stars can potentially be observed by up to 24 cameras, the simulator generates the resulting signals and light curve, for each camera individually. We include all main effects that are currently implemented in \platosim, including the photometry module. At the time when the dataset was generated, only on-board algorithms were implemented by this module. This means that once the full processing chain is complete, the flux is extracted at pixel level using optimal aperture photometry \citep[see][]{MarchioriAl_2019}. We underline that the photometry for bright stars will eventually be derived by PSF fitting. This prevents us from making use of centroids for the targets. However, since we have control over the simulation, this is a great baseline to evaluate the performance of the method. To reduce the computational cost, the simulations were not performed on a complete PLATO field of view (i.e. simulating full-frame CCD images) but star per star, on a CCD subfield of $10\times10$~pixels. We also choose to reduce the cadence from the nominal 25~sec to 1~min and limit the simulations to the first four quarters, to cover a one year time span. We underline our objective was not to assess the performances of the instrument but to test the ability of our software to detect transit-like events. Depending on the type of simulation (planetary transit, eclipsing binary, number of contaminants...), and the version of the simulator, the computation time for a single target, on one quarter and for one camera, takes on the order of $\simeq$ 12 minutes for version 3.6 of \platosim{}, while previous versions took around $\simeq$ 7 minutes. Finally, still in an effort to save computational resource, we decided to adapt the simulation to the orbital period of the transiting body, and did not generate light curves on a given quarter when no transit is expected to occur. As a result, the number of quarters for a given star and a given astrophysical signal is not constant, but is tailored to the orbital period of the transit signal.

Table~\ref{tab:totsimus} gives the summary of the number of the different astrophysical signals that were used in this study. Taking into account the fact that simulations cover one year time span, and that we forego quarters where no transit is present, we end up with a total of 16\,094 quarters that were treated as independent light curves. This is all the more relevant as the periodic nature of transits does not come into play in our approach, and that the light curves were not corrected from any trend, such as instrument aging, or even cosmic impacts. We show the resulting distribution of radii and period of the transiting bodies simulated in Fig.~\ref{fig:data_hist}. 

\begin{table}
\centering
\caption{Number of simulations per type of signal injected.}
\begin{tabular}{ | r || r | r | }
    \hline
    Event type & Simulated & Truncated \\
    \hline\hline
    Planetary transit & 3\,593 & 2\,677 \\
    Eclipsing binaries & 2\,005 & 1\,984 \\
    Background eclipsing binary & 741 & 736 \\
    \hline
    Total & 6\,339 & 5\,397 \\
    \hline
\end{tabular}
\label{tab:totsimus}
\end{table}

We further filter the dataset by removing edge cases where, due to numerical errors, the transits were not visible in the quarters. We also truncate the dataset to remove cases where non physical parameters were used to generate light curves. For instance, we remove cases where the stellar radius >2.5\,\rsun, or where the transit depth is <50\,ppm. This leaves us with 14\,594 light curves, that we randomly split into two datasets; 85\% training and 15\% validation, that is 12\,405 and 2\,189 quarters, respectively. The final counts of signals in the dataset used is shown in the right column of Table~\ref{tab:totsimus}.

\section{Performances}\label{sec:performances}

Evaluation of the performance of the model can be done in two ways. First, directly evaluating the raw output compared to the desired label. Second, assessing the ability of the model to detect transit events, or lack thereof. The former is achieved by computing conventional metrics, such as precision, recall, average precision, $F_1$ score and Jaccard score (or intersection over union; IOU). The later is done by comparing the positions of the ground truths of the events to the predicted positions by the model. While the direct approach allows a straight-forward evaluation of the model, it also doesn't reflect its actual ability to detect transits. We therefore focus our estimates on the ability of the model to recover transits, as well as its false positive rate (FAR). We deem a transit to be successfully recovered if an overlap between the prediction and the ground truth exists. The FAR is defined as the fraction of false positives to the total number of predicted event.

Models are trained on A40 GPUs in the in-house cluster of the laboratory. We trained a total of 16 models using the Unet3+ architecture, expected to perform the best. We test multiple initial kernel lengths and trainable parameters, using 4 or 8 initial features, and did this up to 70 training epochs, using batches of 40 light curves per training pass. To find the best performing versions of the models, we check the recovery and FAR performance on the last 20 epochs. We explored two options: a conservative approach that takes the model that has the lowest FAR at the 0.95 confidence threshold, and finding the version of the models that retrieve the most planets for a constant FAR of 1\%. We show the summary of the parameters, and their associated results, in Table~\ref{tab:models_tested}.

When taking the models in their conservative regime, we find that our models are able to retrieve more than 80\% of our test population, and that with a FAR under 0.1\%, less than 1 false positive for 1\,000 predictions. When fixing the FAR to 1\%, we find that we are able to retrieve 90\% of the planets in our test dataset. These performances demonstrate that this approach is not only viable, but beneficial. The inference mechanism is very fast ($\sim0.2$ seconds per light curve on a CPU), allowing for processing large amounts of data, which will be the case of PLATO. In this case, keeping the number of false positives small is key to enabling rapid and accurate processing of the vast amount of light curves.

\begin{table*}[]
    \centering
    \caption{Parameters and performances of every models.}
    \label{tab:models_tested}
    \begin{tabular}{ | c | r r r || c c c | c c | c | }
    \hline
    \multirow{2}{*}{$N_f$} & \multicolumn{3}{ | c || }{Kernel} & \multicolumn{3}{c | }{Conservative} & \multicolumn{2}{c | }{@ 1\% FAR} & \multirow{2}{*}{Label} \\
     & $k_s$ & $k_t$ & d & FAR [\%] & Retrieval [\%] & Epoch & Retrieval [\%] & Epoch & \\
    \hline\hline
    \multirow{8}{*}{4} & 31 & 11 & 3 & 0.02 & 81.64 & 56 & 86.23 & 53 & \\
     & 31 & 31 & 1 & \textbf{<0.01} & 67.34 & 66 & 80.58 & 66 & \\
    \cline{2-10}
     & 61 & 11 & 6 & 0.03 & 84.63 & 57 & 88.26 & 63 & \\
     & 61 & 31 & 2 & \textbf{<0.01} & 82.60 & 61 & 88.15 & 66 & \\
    \cline{2-10}
     & 121 & 11 & 12 & 0.37 & 86.45 & 67 & 89.54 & 52 & \\
     & 121 & 31 & 4 & 0.05 & 81.22 & 70 & 86.34 & 56 & \\
    \cline{2-10}
     & 181 & 11 & 18 & 0.17 & 86.23 & 56 & \textbf{90.07} & \textbf{68} & \textbf{A} \\
     & 181 & 31 & 6 & 0.09 & 86.34 & 69 & 89.22 & 63 & \\
    \hline\hline
    \multirow{8}{*}{8} & 31 & 11 & 3 & 0.02 & 78.76 & 69 & 86.77 & 67 & \\
     & 31 & 31 & 1 & 0.04 & 82.60 & 54 & 86.66 & 53 & \\
    \cline{2-10}
     & 61 & 11 & 6 & \textbf{<0.01} & \textbf{85.81} & \textbf{64} & \textbf{89.64} & \textbf{53} & \textbf{B, C} \\
     & 61 & 31 & 2 & 0.02 & 84.20 & 57 & 89.43 & 66 & \\
    \cline{2-10}
     & 121 & 11 & 12 & 0.09 & 86.65 & 58 & 89.54 & 61 & \\
     & 121 & 31 & 4 & 0.02 & 84.95 & 63 & 88.26 & 64 & \\
    \cline{2-10}
     & 181 & 11 & 18 & 0.17 & 87.19 & 63 & 89.43 & 69 & \\
     & 181 & 31 & 6 & 0.03 & 84.52 & 52 & 89.43 & 69 & \\
    \hline
    \end{tabular}
    \tablefoot{Performances of the models tested for this study. We consider two approaches to evaluate the results of the models, and compare them for each initial number of features $N_f$, kernel size $k_s$ and trainable parameters in the kernel $k_t$. The conservative approach presents the model with the lowest FAR possible, and the second consists in fixing the FAR at a value of 1\%, and taking the highest resulting recovery.}
\end{table*}

\begin{figure}
	\centering
	\includegraphics[width = \columnwidth]{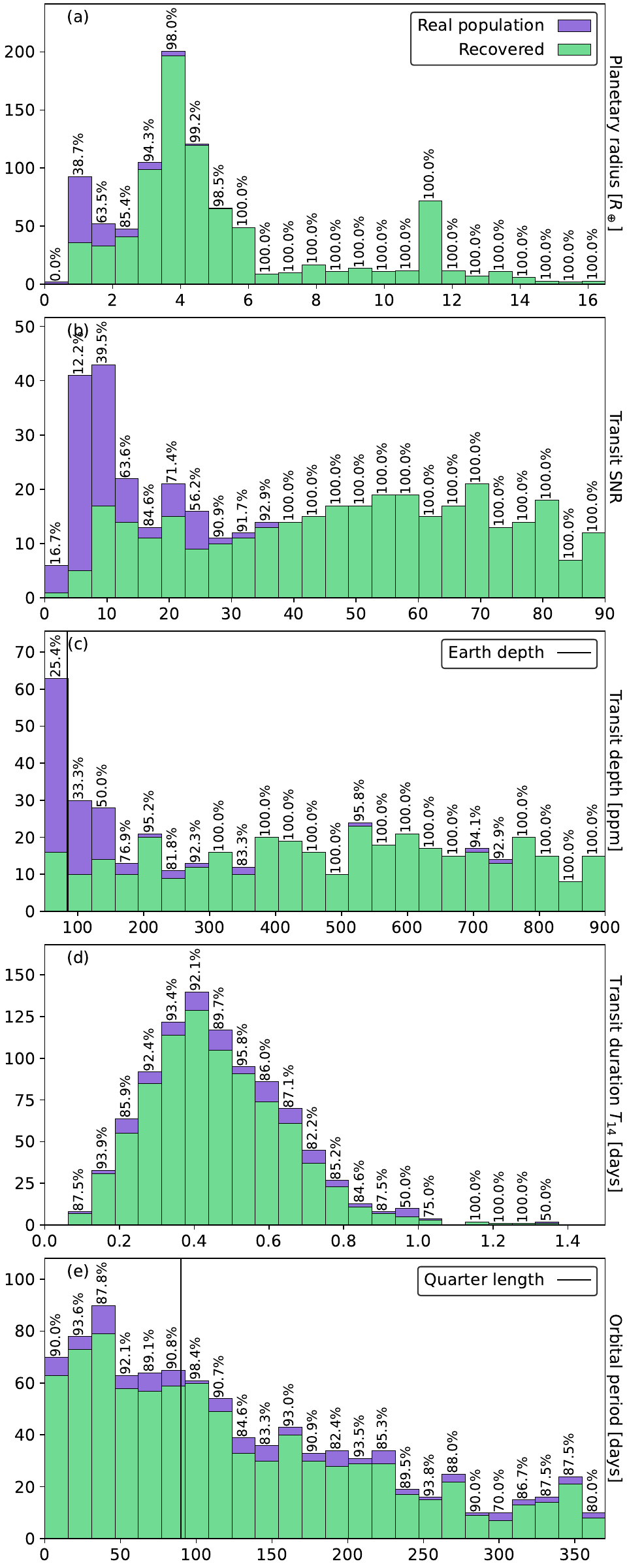}
	\caption{The recovery capabilities of model C. Each panel shows a physical characteristic of our test population, in purple, as well as the fraction recovered by the model, in green. Panel c highlights the ability of the model to recover transits similar to that of Earth. Additionally, panel e shows that the ability to detect transits is not linked to the orbital period, and single transit are therefore detected.}
	\label{fig:recovery}
\end{figure}

As highlighted in Table~\ref{tab:models_tested}, we consider three models that offer the best performances. Model A retrieves the largest fraction of the test population, model B yields a FAR of less than 0.01\% while still recovering more than 85\% of the planets, and finally model C provides a solid compromise between recovery and FAR. We use these models as an illustration for our approach on the population of our dataset. We show in Fig.~\ref{fig:recovery} the effectiveness of model C at 1\% FAR at recovering the planets in our test population. The limiting factor for detection that emerges is the depth of the transits, and their associated SNR. We here compute the SNR after \citet{HowardAl_2012}:

\begin{equation}\label{eq:snr}
    \text{SNR} = \frac{\delta}{\sigma_\text{CDPP}}\sqrt{\frac{n_\text{tr}\cdot t_\text{dur}}{\text{3hr}}}
\end{equation}

\noindent where $\delta$ is the depth of the transit, $\sigma_\text{CDPP}$ is the combined differential photometric precision, $n_\text{tr}$ is the number of observed transits and $t_\text{dur}$ the transit duration. While the recovery rate noticeably drops for depths lower than $\sim150$\,ppm (SNR of $\sim15$), Earth-analogs are detectable by the model. Fig.~\ref{fig:recovery} (c) shows the depths of the events where the expected Earth depth is highlighted as a black vertical line, and neighboring planets are recovered at a rate between 25--33\%. Additionally, the duration of transits are found to have little impact on the recovery rate (panel d), and crucially, the orbital period also has no impact on transit recovery (panel e). This holds true even for planets with orbital periods longer than a single quarter, indicating that transits are indeed identified on a unique event basis. We therefore find that our approach should be be able to identify at least 25\% of the Earth-analogs robustly.

We also subsequently train the Unet and Unet++ architectures to compare their performances relative to the best Unet3+ versions. We find that these alternative models perform slightly worse. Namely, we find that the recovery is lower at equal FAR, especially in the small planets regime. We therefore limit our analysis to the pest-performing Unet3+ models.

\begin{figure}
    \centering
    \includegraphics[width = \columnwidth]{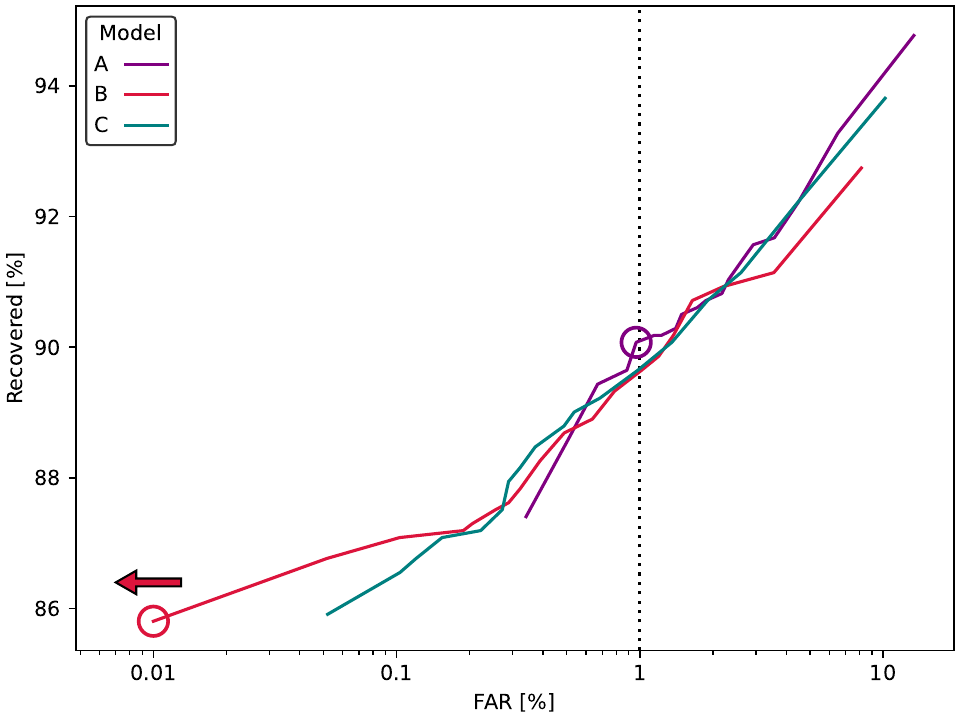} \\
    \includegraphics[width = \columnwidth]{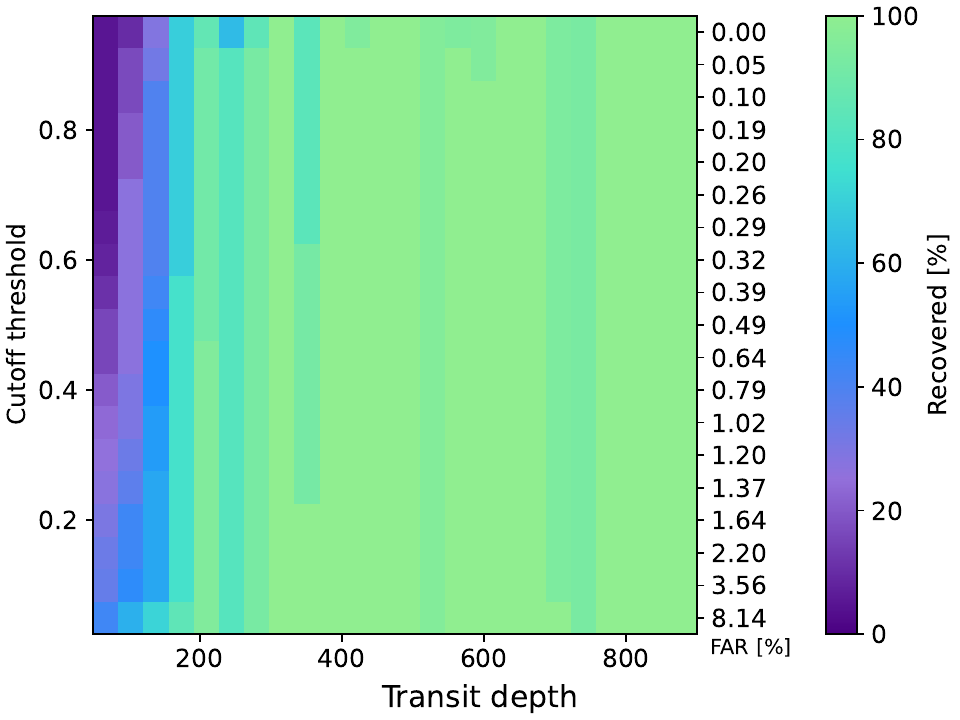}
    \caption{Recovery and FAR for the best performing models. The top panel shows the receiver operating characteristic curve, that links the recovery percentage to its associated FAR. We show the epoch corresponding to each selected models, and mark the recovery/FAR balance highlighted in Table~\ref{tab:models_tested}. The bottom panel shows model B, binned per transit depth. By discretizing the recovery, we can evaluate the performances of the model more thoroughly.}
    \label{fig:last}
\end{figure}

To better highlight the performances of the models, we illustrate in the top panel of Fig.~\ref{fig:last} the trade-off between the recovery rate and the FAR in a receiver operating characteristic (ROC). We show the three selected models and their compromise, identifying the FAR selected in Table~\ref{tab:models_tested}. We see for each case that the number of planets recovered increases with the FAR. Importantly, we find that even for the lowest possible FAR, here model B with $<0.01\%$, a sizable 85.81\% of the population is successfully recovered. The lower panel of Fig.~\ref{fig:last} illustrates the recovery for various depths of transits. It illustrates clearly that the recovery rate rapidly rises above $\sim180$\,ppm, essentially reaching 100\% (as also visible in panel c of Fig.~\ref{fig:recovery}, for model C).

\begin{figure*}
    \centering
    \includegraphics[width = \textwidth, draft = false]{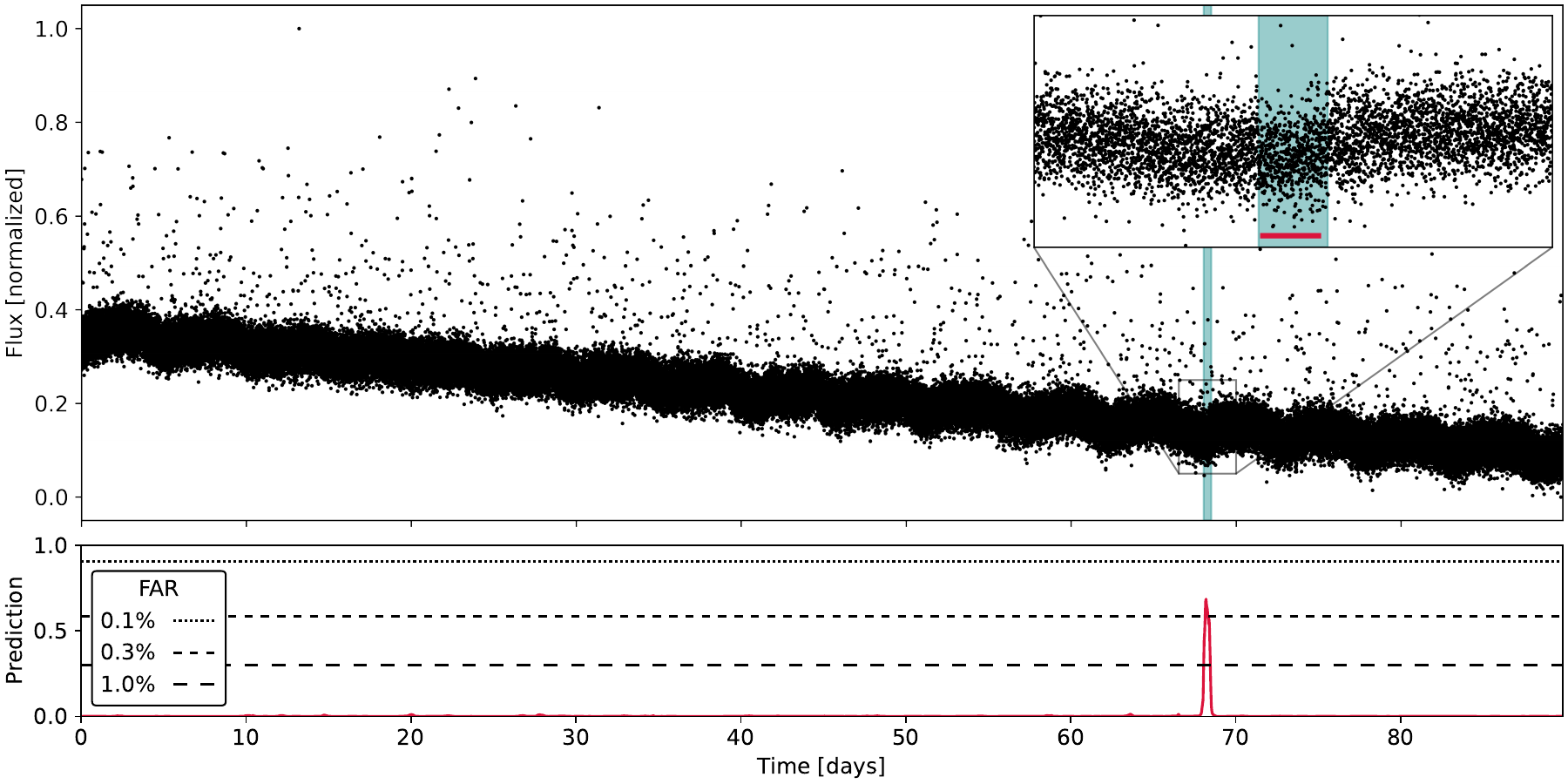}
    \caption{Detection example of an Earth-analog with a depth of 83.16\,ppm (SNR of 11.69), using model C. The top panel shows the light curve, with the ground truth of the transit is shown as a blue span. The bottom panel shows the associated prediction map of the model, with the associated FARs of the model shown as dashed lines. The zoomed in section highlights the transit, and overlays the predicted position of that transit by the model as a red line, extracted at the 1\% FAR.}
    \label{fig:detection}
\end{figure*}

We illustrate the detection of an Earth-analog signal ($R_P=1.11\,R_\oplus$, $R_S=1.23\,R_\odot$, $\delta=83.16$\,ppm) in Fig.~\ref{fig:detection}, using model C. This unique event is recovered with a confidence level of more than 0.65, making this planet detected at a corresponding FAR of less than 0.3\%. While not strictly equivalent to the false alarm probability, it sufficiently analogous to give insight on the likelihood that this event is a true positive.

\section{Discussion}\label{sec:discussion}

In this paper, we presented \panopticon{}, a DL approach designed to detect unique transit event in unfiltered PLATO light curves. We trained 16 versions of the models using various hyperparameters to test the robustness of the method and find the best performing iterations. We featured three versions of the model, corresponding best versions for recovery rate (A), lowest FAR (B), and best trade-off between recovery and FAR (C). We trained and tested our method on simulated PLATO light curves, generated using the \texttt{Platosim} package. Our dataset is made up of a total of 14\,594 light curves, which is split 85\%/15\% into training/testing subsets.

By fixing the FAR at 1\%, we are able to retrieve 90\% of the test population in our dataset. Reciprocally, for the model with the lowest FAR, $<0.01$\%, the recovery rate is still of 85.81\%. Finally, the model presenting the best mixed characteristics is able to retrieve 89.64\% of the population at 1\% FAR, and 85.91\% retrieval at 0.05\% FAR. We find that the only limiting factor in detection is the apparent depth (and subsequent SNR), while neither the duration of the event nor their period that prevent detection. This means that unique transits are indeed recovered successfully, without required any prior detrending. Additionally, we note that Earth-analog signals, that is, unique and shallow transits, are also recovered at a rate slightly greater than 25\%. Any signals resulting in a depth greater than 180\,ppm are almost systematically recovered, even at a low FAR.

The ability of the model to work without prior filtering of the data significantly simplifies the process of finding, and investigating, TCEs. For example, by avoiding drowning small signals during filtering. Also, with the detection being based on single events, the model is able to reliably detect long period exoplanets that might only appear once in a dataset. Computation time for detection is essentially instantaneous, making it a very easy tool to use. This is achieved with a training time of around 10 hours per model, using our training dataset of 12\,405 light curves.

We have here successfully applied our approach to PLATO data, and work is currently underway to adapt it to TESS light curves. This will enable the characterization of the impact of real world data rather than simulated. To further the capacity of the model, it is possible to develop a approach coupling our model in high-FAR, high-recovery regime to a vetting model. This would allow a fair increase of the recall, without increasing the FAR.

Additionally, when PLATO becomes operational, using the best performing model as a baseline to train a dedicated model incorporating annotated real-world data will not only be efficient, but ensure a solid foundation.


\begin{acknowledgements}
      HV and MD acknowledge funding from the Institut Universitaire de France (IUF) that made this work possible. This research made use of the computing facilities operated by the CeSAM data center at the LAM, Marseille, France.
\end{acknowledgements}

%
%
\bibliographystyle{aa}
\bibliography{
    bib/science.bib,
    bib/tools.bib,
    bib/ml_astro.bib,
    bib/ml_models.bib,
    bib/ml_other.bib
}


\begin{appendix}
\section{Model architectures}\label{app:architectures}


\tikzset{
    backbone/.style = {
        draw = black,
        very thick
    },
    neuron/.style = {               
        circle,
        draw = black!80,
        fill = black!10,
        very thick,
        minimum size = 1cm,
        node distance = 0.75cm and 0.75cm
    },
    convolution/.style = {          
        arrows = {->[color=RoyalPurple]},
        draw = RoyalPurple,
        line width = 0.8mm
    },
    upconv/.style = {               
        arrows = {->[color=YellowOrange]},
        draw = YellowOrange,
        line width = 0.8mm
    },
    skip/.style = {                 
        arrows = {->[color=Emerald]},
        draw = Emerald,
        dashed,
        line width = 0.45mm
    },
    nonlin/.style = {               
        arrows = {->[color=black]},
        draw = black,
        line width = 0.8mm
    }
}

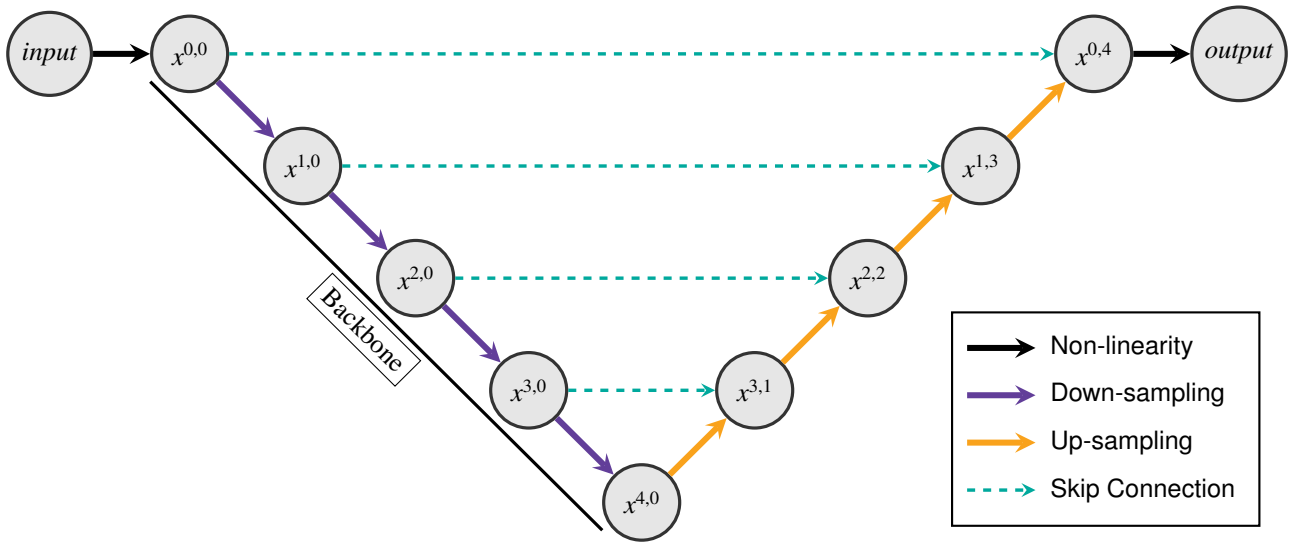
\begin{figure*}[h]
\centering
    \begin{tikzpicture}[x=1.5cm, y=1.5cm, >=stealth]
    
        \node[neuron](input){$\mathit{input}$};
        \node[neuron](x00)[right = of input]{$x^{0,0}$};
        \node[neuron](x10)[below right = of x00]{$x^{1,0}$};
        \node[neuron](x20)[below right = of x10]{$x^{2,0}$};
        \node[neuron](x30)[below right = of x20]{$x^{3,0}$};
        \node[neuron](x40)[below right = of x30]{$x^{4,0}$};
    
        \node[neuron](x31)[above right = of x40]{$x^{3,1}$};
        \node[neuron](x22)[above right = of x31]{$x^{2,2}$};
        \node[neuron](x13)[above right = of x22]{$x^{1,3}$};
        \node[neuron](x04)[above right = of x13]{$x^{0,4}$};
        \node[neuron](output)[right = of x04]{$\mathit{output}$};
    
        \path[convolution]  (x00) edge (x10)
                            (x10) edge (x20)
                            (x20) edge (x30)
                            (x30) edge (x40);
        
        \path[upconv]       (x40) edge (x31)
                            (x31) edge (x22)
                            (x22) edge (x13)
                            (x13) edge (x04);
        
        \path[skip]         (x00) edge (x04)
                            (x10) edge (x13)
                            (x20) edge (x22)
                            (x30) edge (x31);
        
        \path[nonlin]       (input) edge (x00)
                            (x04) edge (output);

        \path[backbone] ($(x00.south west) + (-0.1, 0)$) edge ($(x40.south west) + (-0.1, 0)$);
        \node[draw, rotate=-45] at ($(x20.south west) + (-0.21,-0.21)$) {Backbone};
        
        \path ([xshift=4cm, yshift=-0.5cm]current bounding box.center)
            node[
                matrix,
                anchor = north west,
                cells = {nodes = {font = \sffamily, anchor = west}},
                draw,
                thick,
                inner sep=1ex
            ]{
                \draw[nonlin](0,0) -- ++ (0.6,0); & \node{Non-linearity};\\
                \draw[convolution](0,0) -- ++ (0.6,0); & \node{Down-sampling};\\
                \draw[upconv](0,0) -- ++ (0.6,0); & \node{Up-sampling};\\
                \draw[skip](0,0) -- ++ (0.6,0); & \node{Skip Connection};\\
        };
    
    \end{tikzpicture}
\caption{The Unet architecture, presented here with depth of 4. The architecture makes use of down- and up-sampling steps of the original input. This allows the extraction of features of various sizes, and recombining them during decoding via the plain skip connections.}
\label{graph:unet}
\end{figure*}


\begin{figure*}[h]
\centering
    \begin{tikzpicture}[x=1.5cm, y=1.5cm, >=stealth]
    
        \node[neuron](input){$\mathit{input}$};
        \node[neuron](x00)[right = of input]{$x^{0,0}$};
        \node[neuron](x10)[below right = of x00]{$x^{1,0}$};
        \node[neuron](x20)[below right = of x10]{$x^{2,0}$};
        \node[neuron](x30)[below right = of x20]{$x^{3,0}$};
        \node[neuron](x40)[below right = of x30]{$x^{4,0}$};
    
        \node[neuron](x31)[above right = of x40]{$x^{3,1}$};
        \node[neuron](x22)[above right = of x31]{$x^{2,2}$};
        \node[neuron](x13)[above right = of x22]{$x^{1,3}$};
        \node[neuron](x04)[above right = of x13]{$x^{0,4}$};
        \node[neuron](output)[right = of x04]{$\mathit{output}$};
    
        \node[neuron](x01)[above right = of x10]{$x^{0,1}$};
        \node[neuron](x11)[above right = of x20]{$x^{1,1}$};
        \node[neuron](x02)[above right = of x11]{$x^{0,2}$};
        \node[neuron](x21)[above right = of x30]{$x^{2,1}$};
        \node[neuron](x12)[above right = of x21]{$x^{1,2}$};
        \node[neuron](x03)[above right = of x12]{$x^{0,3}$};
    
        \path[convolution]  (x00) edge (x10)
                            (x10) edge (x20)
                            (x20) edge (x30)
                            (x30) edge (x40);
        
        \path[upconv]       (x30) edge (x21)
                            (x21) edge (x12)
                            (x12) edge (x03)
                            (x10) edge (x01)
                            (x20) edge (x11)
                            (x11) edge (x02)
                            (x40) edge (x31)
                            (x31) edge (x22)
                            (x22) edge (x13)
                            (x13) edge (x04);
        
        \path[skip]         (x00) edge (x01)
                            (x00) edge[bend left = 18] (x02)
                            (x00) edge[bend left = 18] (x03)
                            (x00) edge[bend left = 20] (x04)
                            (x01) edge (x02)
                            (x01) edge[bend left = 18] (x03)
                            (x01) edge[bend left = 18] (x04)
                            (x02) edge (x03)
                            (x02) edge[bend left = 18] (x04)
                            (x03) edge (x04)
                            (x10) edge (x11)
                            (x10) edge[bend left = 18] (x12)
                            (x10) edge[bend left = 18] (x13)
                            (x11) edge (x12)
                            (x11) edge[bend left = 18] (x13)
                            (x12) edge (x13)
                            (x20) edge (x21)
                            (x20) edge[bend left = 18] (x22)
                            (x21) edge (x22)
                            (x30) edge (x31);
        
        \path[nonlin]       (input) edge (x00)
                            (x04) edge (output);

        \path[backbone] ($(x00.south west) + (-0.1, 0)$) edge ($(x40.south west) + (-0.1, 0)$);
        \node[draw, rotate=-45] at ($(x20.south west) + (-0.21,-0.21)$) {Backbone};
        
        \path ([xshift=4cm, yshift=-1cm]current bounding box.center)
            node[
                matrix,
                anchor = north west,
                cells = {nodes = {font = \sffamily, anchor = west}},
                draw,
                thick,
                inner sep=1ex
            ]{
                \draw[nonlin](0,0) -- ++ (0.6,0); & \node{Non-linearity};\\
                \draw[convolution](0,0) -- ++ (0.6,0); & \node{Down-sampling};\\
                \draw[upconv](0,0) -- ++ (0.6,0); & \node{Up-sampling};\\
                \draw[skip](0,0) -- ++ (0.6,0); & \node{Skip Connection};\\
        };
    
    \end{tikzpicture}
\caption{The Unet++ architecture of same depth as Fig.~\ref{graph:unet}. This version introduces a more complex recombination during decoding. Each encoding level is up-sampled individually and combined in nested dense skip connections. This gives a better merging of various feature sizes when creating the output.}
\label{graph:unetpp}
\end{figure*}
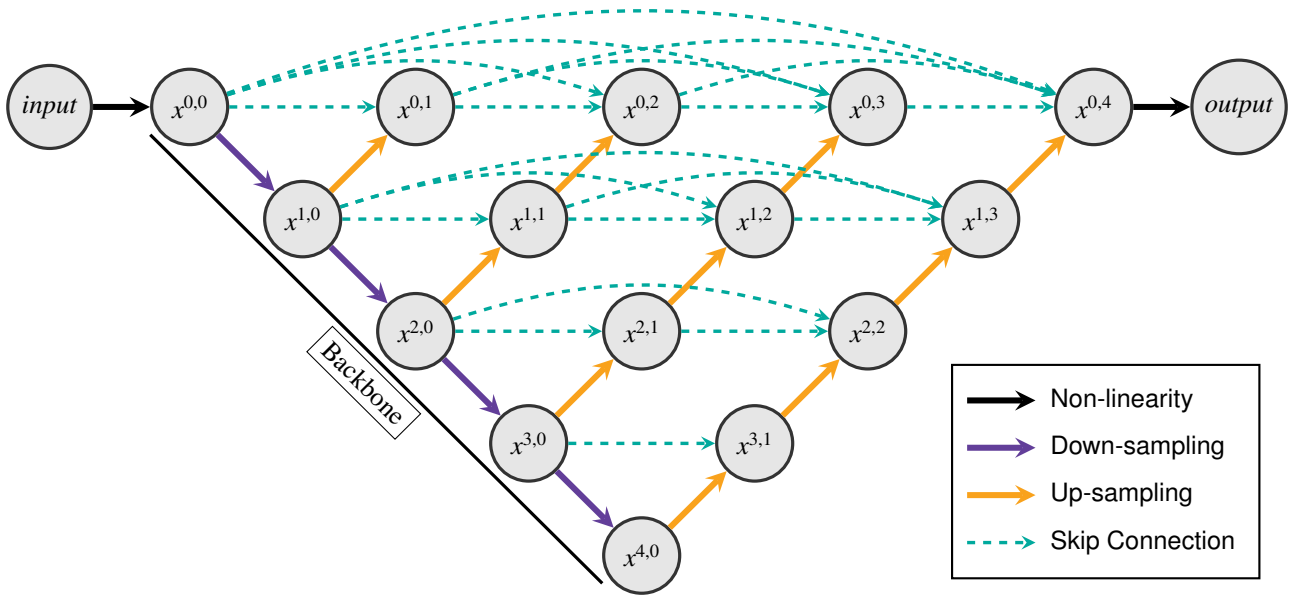


\tikzset{
    skipex/.style = {               
        arrows = {->[color=black!60]},
        draw = black!60,
        dashed,
        line width = 0.45mm
    },
    skip04/.style = {               
        arrows = {->[color=YellowOrange]},
        draw = YellowOrange,
        dashed,
        line width = 0.45mm
    },
    skip13/.style = {               
        arrows = {->[color=YellowGreen]},
        draw = YellowGreen,
        dashed,
        line width = 0.45mm
    },
    skip22/.style = {               
        arrows = {->[color=Emerald]},
        draw = Emerald,
        dashed,
        line width = 0.45mm
    },
    skip31/.style = {               
        arrows = {->[Mulberry]},
        draw = Mulberry,
        dashed,
        line width = 0.45mm
    }
}

\begin{figure*}[h]
\centering
    \begin{tikzpicture}[x=1.5cm, y=1.5cm, >=stealth]
    
        \node[neuron](input){$\mathit{input}$};
        \node[neuron](x00)[right = of input]{$x^{0,0}$};
        \node[neuron](x10)[below right = of x00]{$x^{1,0}$};
        \node[neuron](x20)[below right = of x10]{$x^{2,0}$};
        \node[neuron](x30)[below right = of x20]{$x^{3,0}$};
        \node[neuron](x40)[below right = of x30]{$x^{4,0}$};
    
        \node[neuron](x31)[above right = of x40]{$x^{3,1}$};
        \node[neuron](x22)[above right = of x31]{$x^{2,2}$};
        \node[neuron](x13)[above right = of x22]{$x^{1,3}$};
        \node[neuron](x04)[above right = of x13]{$x^{0,4}$};
        \node[neuron](output)[right = of x04]{$\mathit{output}$};
    
        \path[convolution]  (x00) edge (x10)
                            (x10) edge (x20)
                            (x20) edge (x30)
                            (x30) edge (x40);
        
        \path[skip04]       (x00) edge (x04)
                            (x13) edge (x04)
                            (x22) edge[bend right = 25] (x04)
                            (x31) edge[bend right = 25] (x04)
                            (x40) edge[bend right = 25] (x04);

        \path[skip13]       (x00) edge (x13)
                            (x10) edge (x13)
                            (x22) edge (x13)
                            (x31) edge[bend right = 25] (x13)
                            (x40) edge[bend right = 25] (x13);
                            
        \path[skip22]       (x00) edge (x22)
                            (x10) edge (x22)
                            (x20) edge (x22)
                            (x31) edge (x22)
                            (x40) edge[bend right = 25] (x22);

        \path[skip31]       (x00) edge (x31)
                            (x10) edge (x31)
                            (x20) edge (x31)
                            (x30) edge (x31)
                            (x40) edge (x31);

        \path[nonlin]       (input) edge (x00)
                            (x04) edge (output);

        \path[backbone] ($(x00.south west) + (-0.1, 0)$) edge ($(x40.south west) + (-0.1, 0)$);
        \node[draw, rotate=-45] at ($(x20.south west) + (-0.21,-0.21)$) {Backbone};

        \path ([xshift=3.7cm, yshift=-1cm]current bounding box.center)
            node[
                matrix,
                anchor = north west,
                cells = {nodes = {font = \sffamily, anchor = west}},
                draw,
                thick,
                inner sep=1ex
            ]{
                \draw[nonlin](0,0) -- ++ (0.6,0); & \node{Non-linearity};\\
                \draw[convolution](0,0) -- ++ (0.6,0); & \node{Down-sampling};\\
                \draw[skipex](0,0) -- ++ (0.6,0); & \node{Full Skip Connection};\\
        };
    
    \end{tikzpicture}
\caption{The Unet3+ architecture of same depth as Figs.~\ref{graph:unet}\&\ref{graph:unetpp}. The skip connections are here not up-sampled for each encoder level, but are included directly when computing the decoder levels, and merged with previous decoder levels. This creates a simpler decoding process, limiting the number of free parameters compared to Unet++, as there are no in between convolution layer.}
\label{graph:unet3p}
\end{figure*}
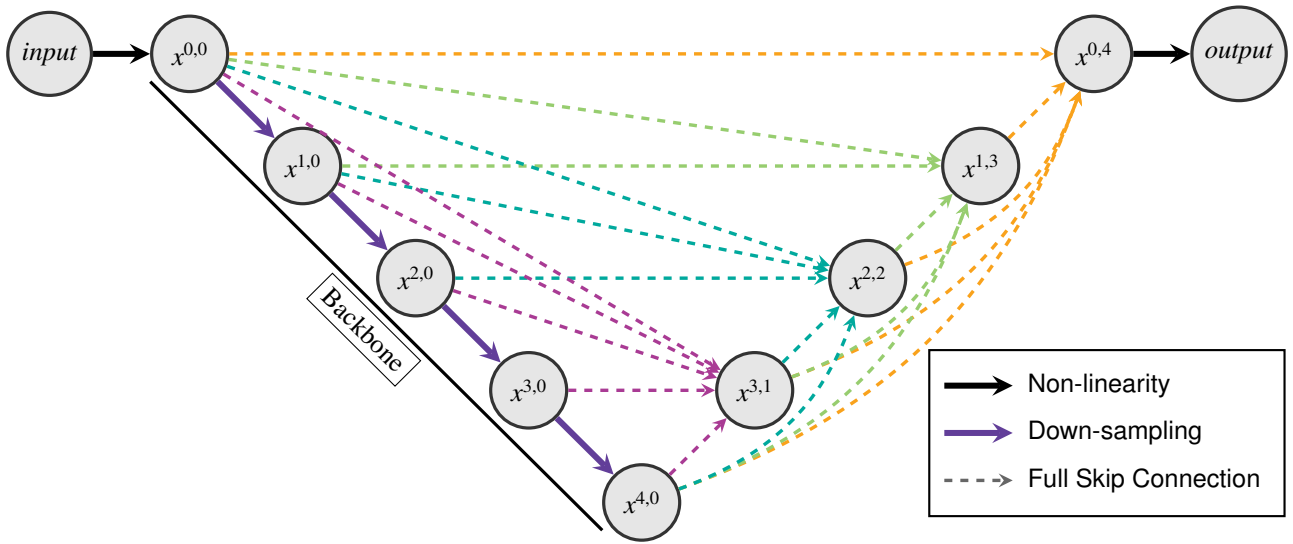

\end{appendix}
\end{document}